\documentclass[jkps,preprint,fleqn,showpacs,showkeys]{revtex4}
\usepackage{graphicx}
\usepackage{amssymb}
\usepackage{amsmath}
\usepackage{bm}
\begin{document}
\setcounter{page}{0}
\title[]{Electronic phase separation due to magnetic polaron formation in the semimetallic ferromagnet EuB$_6$ --\\ A weakly-nonlinear-transport study}
\author{Adham \surname{Amyan}}
\author{Pintu \surname{Das}}
\author{Jens \surname{M\"uller}}
\email{j.mueller@physik.uni-frankfurt.de}
\thanks{Fax: +49-69-798-47227}
\affiliation{Institute of Physics, Goethe-University Frankfurt, 60438 Frankfurt (M), Germany}
\author{Zachary \surname{Fisk}}
\affiliation{Department of Physics, University of California, Irvine, California 92697, USA}

\date[]{Received \today}

\begin{abstract}
We report measurements of weakly nonlinear electronic transport, as measured by third-harmonic voltage generation $V_{3\omega}$, in the low-carrier density semimetallic ferromagnet EuB$_6$, which exhibits an unusual magnetic ordering with two consecutive transitions at $T_{c_1} = 15.6$\,K and  $T_{c_2} = 12.5$\,K. In contrast to the linear resistivity, the third-harmonic voltage is sensitive to the microgeometry of the electronic system. Our measurements provide evidence for magnetically-driven electronic phase separation consistent with the picture of percolation of magnetic polarons (MP), which form highly conducting magnetically ordered clusters in a paramagnetic and less conducting background. Upon cooling in zero magnetic field through the ferromagnetic transition, the dramatic drop in the linear resistivity at the upper transition $T_{c_1}$ coincides with the onset of nonlinearity, and upon further cooling is followed by a pronounced peak in $V_{3 \omega}$ at the lower transition $T_{c_2}$. Likewise, in the paramagnetic regime, a drop of the material's magnetoresistance $R(H)$ precedes a magnetic-field-induced peak in nonlinear transport. A striking observation is a linear temperature dependence of $V_{3\omega}^{\rm peak}(H)$.
We suggest a picture where at the upper transition $T_{c_1}$ the coalescing MP form a conducting path giving rise to a strong decrease in the resistance. The MP formation sets in at around $T^\ast \sim 35$\,K below which these entities are isolated and strongly fluctuating, while growing in number. The MP then start to form links at $T_{c_1}$, where percolative electronic transport is observed. The MP merge and start forming a continuum at the threshold  $T_{c_2}$. In the paramagnetic temperature regime $T_{c_1} < T < T^\ast$, MP percolation is induced by a magnetic field, and the threshold accompanied by charge carrier delocalization occurs at a single critical magnetization.
\end{abstract}



\maketitle

\section{Introduction}
Spintronics research exploits the fact that the magnetic state of a system critically affects its electronic transport properties. Over the years, a number of more or less complex model systems have been identified showing large negative (and even colossal) magnetoresistance (MR) behavior, among them the rare-earth chalcogenides, diluted and concentrated magnetic semiconductors, and mixed-valence perovskites (manganites). These classes of materials often have rich phase diagrams, in which many phases exhibit intrinsic, i.e.\ non-chemical, electronic phase separation. Nanoscale electronic phase separation in turn is thought to play a critical role in the emergence of such spectacular physical phenomena as high-temperature superconductivity (HTSC) in cuprates and colossal magnetoresistance (CMR) in magnetic semiconductors and manganites. Therefore, electronic phase separation has been a subject of intensive recent theoretical and experimental interest.\\
In order to study some fundamental aspects of electronic phase separation and CMR behavior, we have chosen the  low-carrier-density ferromagnetic semimetal EuB$_6$, which is considered simple because of the cubic lattice symmetry and the absence of magnetocrystalline anisotropy for the Eu$^{2+}$ localized 4$f$ spins in a $^8S_{7/2}$ Hund's rule ground state. Yet, it shows interesting physics, where the interplay of metallicity and the formation of clustered magnetic phases can be studied on the fundamental level.

\section{Physical Properties of EuB$_6$}
Despite the simple lattice and magnetic structure, the physical properties at low temperatures, in particular the mechanism of ferromagnetic ordering in EuB$_6$ and its interplay with the CMR behavior is far from being fully understood. While the material undergoes the paramagnetic (PM) to ferromagnetic (FM) transition, it exhibits two anomalous features at $T_{c_1} \sim 15.5$\,K and $T_{c_2} \sim 12.5$\,K in electronic transport and specific heat measurements, which initially have been interpreted as different kinds of FM ordering \cite{DegiorgiPRL1997,CooleyPRB1997}. Applying small magnetic fields drastically suppresses the resistivity at the higher transition $T_{c_1}$, with a negative MR at 15\,K being larger than 90\,\%. Different mechanisms have been discussed in order to explain the CMR effect in nonmanganite systems, e.g.\ the suppression of critical magnetic fluctuations with externally applied magnetic fields $H$ \cite{MajumdarPRL1998} or a delocalization of carriers due to the overlap of magnetic polarons (MP) \cite{TeresaNature1997,NyhusPRB1997,SuellowPRB2000}. MP are formed when it is energetically favorable for the charge carriers to spin polarize the local moments over a finite distance, i.e.\ the localization length of the charge carrier \cite{KasuyaRMP1968}. 
The size of polaronic clusters of ferromagnetically oriented localized spins around the charge carrier spins is thus determined by the balance of the increase in energy of the charge carriers due to their localization and the reduction of exchange energy due to alignment of the local moments \cite{vonMolnarHandbook}. The motion of an isolated MP or small cluster, i.e.\ hopping of the localized carrier, is impeded by the PM background spins. An external magnetic field or magnetic order aligns the outside spins, thereby strongly reducing the impedance of MP motion. When the MP coalesce, the carriers finally delocalize. From small angle neutron scattering experiments on manganites, De Teresa \textit{et al.}\ have demonstrated the existence of such clustered phases above the ferromagnetic ordering temperature $T_C$ of manganite materials \cite{TeresaNature1997}. Although the formation of MP has been discussed for various different magnetic systems, e.g., dilute and concentrated magnetic semiconductors, rare-earth chalcogenides, manganites, pyrochlores etc., the underlying microscopic nature of electrical transport in the MP phases is not yet properly understood and is a matter of current debate \cite{MajumdarPRL1998, CalderonPRB2004, ChatterjeePRB2004, YuPRB2006}.\\
Accordingly, in EuB$_6$ the transition at $T_{c_1}$ has been ascribed to a metallization transition $T_M$ via the overlap of magnetic polarons \cite{SuellowPRB2000}. Furthermore, the authors argue that the CMR is not associated with the bulk Curie temperature being $T_C = T_{c_2}$ but with $T_M$, and that the separation of the charge delocalization and bulk magnetic ordering transitions implies electronic and magnetic phase separation. Indeed, magnetic phase separation has been detected in muon spin rotation experiments, where two different magnetic environments are found to be important between $T_M = T_{c_1}$ and $T_C = T_{c_2}$, which disappears below $T_{c_2}$ \cite{BrooksPRB2004}. Direct evidence for a magnetically-driven electronic phase separation and percolation from transport measurements, however, only has been observed very recently \cite{Das2012}. An important hint came from recent Hall effect measurements by Zhang {\it et al.}, who observed a distinct change in the slope of the Hall resistivity of EuB$_6$ in the PM phase \cite{ZhangPRL2008,ZhangPRL2009}. 
The switching field in the Hall effect depends linearly on temperature and extrapolates to the paramagnetic Curie temperature of the material, which indicates that the switching in the slope of the Hall voltage occur at a single critical magnetization over a wide temperature range. The authors interpret the critical magnetization as the point of percolation for patches of a more conducting and magnetically ordered phase in a less ordered background. Based on this picture, they obtained excellent scaling and quantitative fits to the Hall effect data through an empirical two-component model \cite{ZhangPRL2009}, providing a measure of the degree of electronic phase separation.

Motivated by these findings, we performed weakly-nonlinear-transport (third-harmonic voltage) measurements, which -- unlike the resistivity itself -- are sensitive to the microgeometry of the sample, i.e.\ intrinsic electronic inhomogeneities. We find  evidence for electronic phase separation and magnetically-driven percolation, at distinct temperatures when cooling through the FM transition and when applying magnetic fields in the PM temperature regime. In the PM temperature regime, a third-harmonic voltage signal is induced by a magnetic field. Similar as for correlated Jahn-Teller polarons in manganites \cite{MoshnyagaPRB2009} the nonlinear signal may reflect the density of MP, which we can describe by a single scaling function for different temperatures and magnetic fields. We obtain a $H$-$T$ phase diagram and suggest a mechanism for the MP behavior at the different temperature scales.

\section{Experiment}
Single crystals of EuB$_6$ were grown from Al flux as described in \cite{FiskJAP1979}.
The linear and nonlinear transport measurements were carried out in standard four-probe geometry using {\it ac} lock-in technique at a frequency of 17\,Hz. For weakly nonlinear transport, current density and electrical field are related by $j = (\sigma  + b|E|^2 + \cdots) E$, where $\sigma$ is the Ohmic or linear conductivity and $b$ the nonlinear conductivity coefficient, with $b|E|^2 \ll \sigma$ \cite{BergmanPRB1989,Levy1994}. The cubic nonlinearity can be accessed in {\it ac} transport measurements by detecting a third-harmonic voltage signal $V_{3\omega}$, see e.g.\ \cite{BergmanPRB1989,DubsonPRB1989}. For a random resistor network with a positive temperature coefficient of the resistance, Dubson {\it et al.}\ have considered the power dissipated in each current-carrying resistor element being proportional to the local temperature oscillation; then, when passing an {\it ac} current $I = I_0 \cos{\omega t}$ through the system, nonlinear transport (as well as the $1/f$ noise power spectral density $S_V$ observed in fluctuation spectroscopy experiments \cite{Das2012}) measures the fourth moment of the current distribution \cite{DubsonPRB1989}. This is because both $V_{3\omega}$ and $1/f$ noise are dominated by the system's weak links through which much of the sample current is channeled. It is 
\begin{equation}
\frac{S_V}{V^2} \propto \frac{V_{3\omega}}{I_0^3R^2} \propto \frac{\sum\limits_\alpha i^4_\alpha}{\left( \sum\limits_\alpha i^2_\alpha \right)^2},
\label{Dubson}
\end{equation}
where $i_\alpha$ are the local currents through the resistor elements $r_\alpha$. In a percolative system close to the threshold, the local current densities and electric fields can be much larger than their average values. This is due to narrow paths, so-called 'bottlenecks' and 'hot spots' in the random network, thus leading to strongly enhanced $1/f$ noise and third-harmonic voltage generation. The direct proportionality Eq.\,(\ref{Dubson}) is fulfilled under certain conditions in classical percolation networks \cite{Bergman1992,YagilPRB1992,DubsonPRB1989}. However, this is not necessarily the case in more complex inhomogeneous systems, as e.g.\ composite systems. In general, however, noise spectroscopy and  nonlinear transport probe the microgeometry of the electronic system, which is not accessible by the linear (Ohmic) resistance, which measures only the second moment of the current distribution. 

\section{Results and Discussion}
\subsection{The ferromagnetic regime}
\begin{figure}
\includegraphics[width=0.66\textwidth]{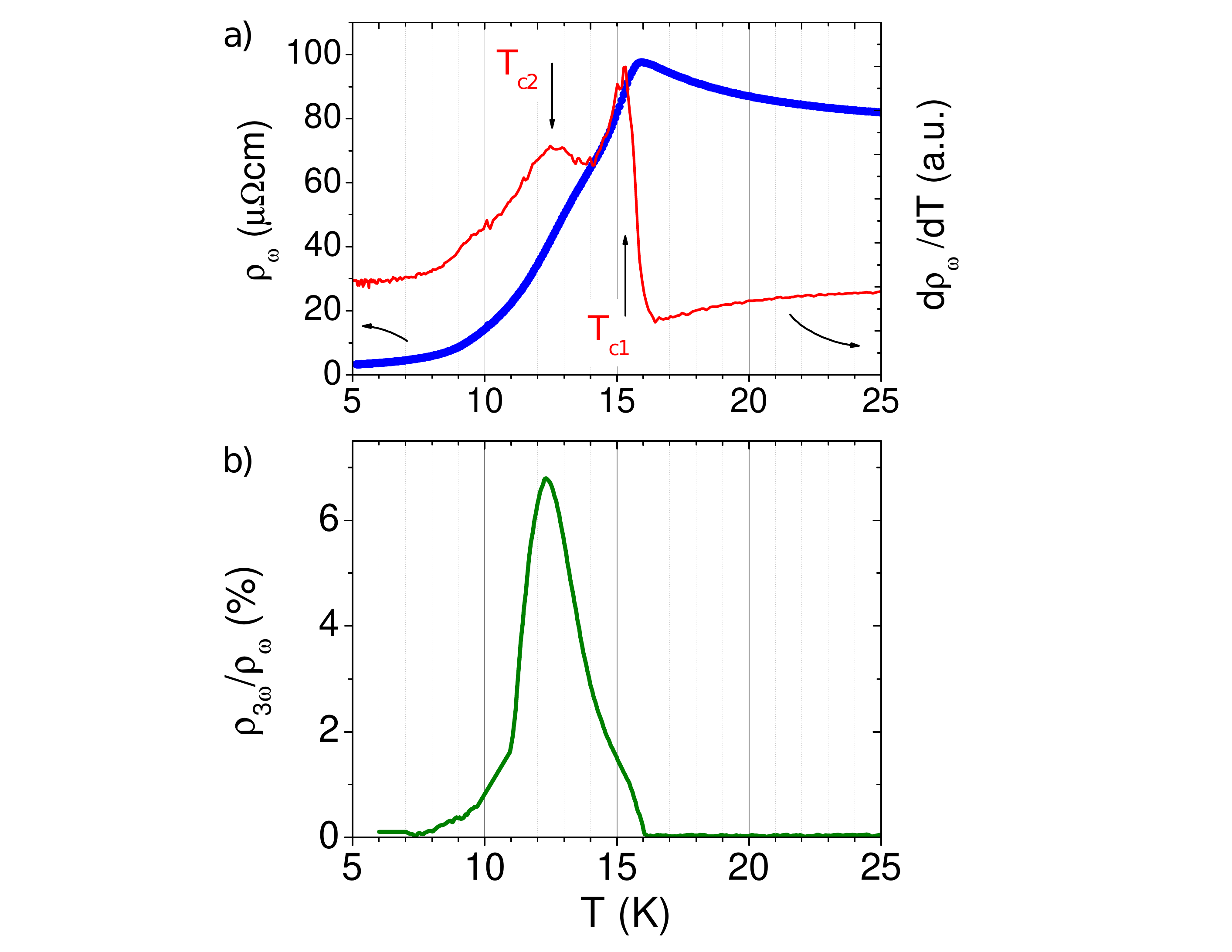}
\caption{(Color online) (a) Resistivity of EuB$_6$ (sample \#1) and temperature derivative ${\rm d}\rho/{\rm d}T$ marking the two transitions at $T_{c_1} = 15.2$\,K and $T_{c_2} = 12.6$\,K. (b) The coefficient of weak nonlinear transport measured by third-harmonic voltage generation, $\kappa = \rho_{3 \omega}/\rho_\omega = V_{3 \omega}/V_\omega$ (sample \#2). One can clearly see the onset of weakly nonlinear transport just above $T_{c_1}$ and a maximum around $T_{c_2}$. Inset shows the linear resistivity in an extended temperature range.
}\label{figure1}
\end{figure}
Figure\,\ref{figure1}(a) shows the bulk linear resistivity of a representative EuB$_6$ sample (\#1). Upon cooling, the resistivity decreases (as expected for a semimetal) from room temperature down to $T^\ast \sim 35$\,K, see inset of Fig.\,\ref{figure1}(b), where a broad minimum is observed. Below that temperature, the resistivity increases and goes through a maximum at about $\sim 16$\,K, which sometimes is used to define the upper transition. Below that temperature, the resistivity rapidly decreases. As seen in Fig.\,\ref{figure1}(a), below that sharp drop there is a shoulder marking the second, lower transition. In the literature, the two transitions are often defined as pronounced peaks in the temperature derivative ${\rm d}\rho/{\rm d}T$ observed here at $T_{c_1} = 15.2 $\,K (just below the resistivity maximum) and $T_{c_2} = 12.6$\,K, see Fig.\,\ref{figure1}(a). Such clearly visible transitions are observed only for high-quality samples \cite{SuellowPRB1998}. This is corroborated by a small value of the residual resistivity $\rho(T \rightarrow 0) \approx 2.3\,\mu\Omega{\rm cm}$ and a large residual resistivity ratio $\rho(300\,{\rm K})/\rho(T \rightarrow 0)$ of about 70, cf.\ Table I.\ in \cite{SuellowPRB1998}. Interestingly, $T_{c_1}$ and $T_{c_2}$ coincide with the onset and peak, respectively, of a weakly nonlinear contribution to the electronic transport, shown in Fig.\,\ref{figure1}(b), measured by the ratio of third-harmonic and linear resistivity $\rho_{3\omega}/\rho_\omega$ shown here for sample \#2. Clearly, in this temperature region, the electronic transport properties are highly inhomogeneous. An explanation of this behavior will be given below. We note that although there is a slight sample-to-sample variation in $T_{c_1}$ and $T_{c_2}$ (16.1\,K and 12.4\,K, respectively for sample \#2, and 15\,K and 12.3\,K for \#3), the relation between the kink in resistance and the peak in the nonlinear signal is ubiquitous in the samples we have measured.\\
The MP are suggested to form (to become stabilized) below $T^\ast \sim 35$\,K \cite{NyhusPRB1997,SuellowPRB2000,BrooksPRB2004}. Below that temperature, the increase in resistance with decreasing temperature may be the result of charge localization by means of magnetic polaron formation \cite{SuellowPRB2000}.  In addition, below the same temperature, a third-harmonic voltage signal can be generated by an external magnetic field, see Fig.\,\ref{figure2} and discussion below.\\
As shown in Fig.\,\ref{figure1}\,(b), the characteristic temperatures $T_{c_1}$ and $T_{c_2}$ are clearly seen also in the weak nonlinear transport, as onset and pronounced peak, respectively. 
In an inhomogeneous material, which may be modeled by a random resistor network or a continuous composite medium, where the local conductivity randomly depends on the radius vector, both weak nonlinear behavior and resistance $1/f$ noise are strongly enhanced as compared to a homogeneous sample with the same characteristics, i.e.\ dimensions and resistance \cite{Kogan1996}. This is due to the strongly nonuniform distribution of current and electric field density, which are confined to narrow paths corresponding to a reduction of the effective 'noisy' volume, which for the inhomogeneous conductor is only a small part of the total volume of the conductor. This effect is extremely pronounced in the vicinity of the percolation threshold, where computer simulations show that the main contribution comes from single bonds, which connect large parts of the infinite cluster, through which a large current passes.
Because both weakly nonlinear transport and $1/f$ noise are sensitive to the fourth moment of the current distribution, Eq.\,(\ref{Dubson}), these quantities are much more sensitive to intrinsic electronic inhomogeneities than the linear resistance, which measures the second moment. In addition, unlike the Ohmic linear resistivity, nonlinear transport and noise are very sensitive to details of the microgeometry like 'hot spots', 'weak links' and 'bottlenecks' resulting in anomalously large local values  $\mathbf{e(r)}$ of the electric field. 

In the picture of formation and percolation of MP, the different temperature regimes marked by characteristic features at $T^\ast$, $T_{c_1}$ and $T_{c_2}$ in resistivity and weakly nonlinear resistivity, as well as $1/f$ noise \cite{Das2012} can be understood if we consider two 'channels' of charge transport: mobile electrons, and holes which are partially localized in the PM regime giving rise to MP formation. When the MP start to form at around $T^\ast \sim 35$\,K, they are isolated and diluted, not forming larger clusters or a network.
Upon cooling, the number and/or size of the MP increases, such that when approaching $T_{c_1}$, the MP form clusters and start to link together. Within coalescing clusters the holes delocalize, which results in the drop of resistance once a continuous conducting path is formed at $T_{c_1}$. This is the percolation threshold $p_c$ for the sample conductance, as reflected by the monotonous increase of effective carrier concentration in this temperature range which saturates below $T_{c_2}$ \cite{PaschenPRB2000,DegiorgiPRL1997}.
Hence, a phase separation scenario applies, where the electronic system can be considered as consisting of magnetically ordered and more conducting spheres, which form clusters when they overlap, and less conducting regions.
When lowering the temperature through $T_{c_1}$ the noise shows a power-law divergence \cite{Das2012}, typical for percolative transport and a finite weakly-nonlinear-transport signal $V_{3 \omega}$ is observed. Both noise and $V_{3 \omega}$ peak at $T_{c_2}$, which in our picture is the percolation threshold $q_c$ for the less conducting regions, which is reached when the MP clusters have mostly merged and start forming a continuum when cooling below $T_{c_2}$, which is also the temperature of bulk magnetic ordering \cite{SuellowPRB2000}. Merging of MP at $T_{c_2}$ leads to a strong reduction of the effective carrier mass at this temperature \cite{PaschenPRB2000}. The weak nonlinear behavior and $1/f$ noise for the present scenario of conducting spheres in a less conducting background, with a variety of possible local microgeometries, have been discussed theoretically \cite{BergmanPRB1989}. It is found that such a composite scenario is more complex than a classical bond or site percolation, because of the role of the local microgeometry --- parameterized by the channel width between the conducting spheres and the degree of their overlap ---, which is related to substantially enhanced local electrical fields.\\ 
Our results highlight the unique electronic and magnetic properties of EuB$_6$ and reveal evidence for magnetically-driven electronic percolation. Close to $T_{c_1}$, the formerly isolated clusters form links, resulting in narrow conducting channels, which then dominate the noise. In this picture, the peak in $V_{3\omega}$ at $T_{c_2}$ reflects the increase of 'bottlenecks' or 'hot spots' in the percolating network of MP, where the local electrical fields $\mathbf{e}$ and current densities $\mathbf{j}$ are strongly enhanced. The contribution of those regions to the overall resistance is proportional to $\mathbf{e(r)^2}$. Since the electrical field dependence of the resistance is given by $R(E) \approx R_\omega + AE^2 + \cdots$ with the second term being the third-harmonic nonlinear resistance, the amplitude of $R_{3\omega}$ indirectly reflects the density and local arrangement of the MP. In the described model, $R_{3\omega}$ is a measure of the configuration of magnetic polaronic clusters, i.e.\ of the microgeometry of the electronic system.

\subsection{The paramagnetic regime}

\begin{figure}
\includegraphics[width=\textwidth]{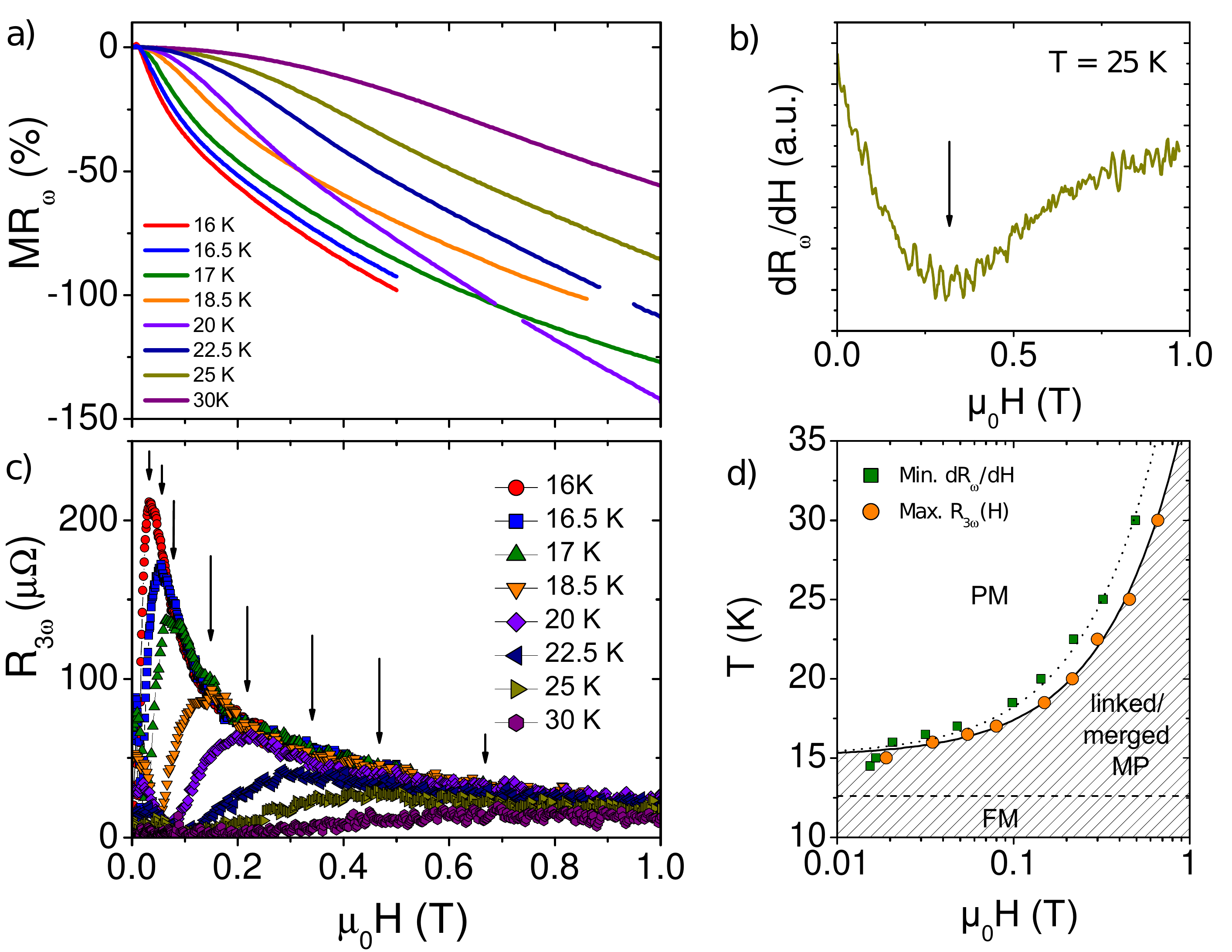}
\caption{ (Color online)(a) Magnetoresistance (MR) of the linear transport, $[R_{\omega}(H) - R_{\omega}(H=0)]/R_{\omega}(H)$, measured at different temperatures in the PM regime (sample \#3). (b) Temperature derivative of the linear MR at $T = 25$\,K exhibiting a pronounced minimum. (c) $R_{3\omega}$ as a function of magnetic field $H$ measured at different temperatures in the PM regime. Arrows indicate the position of pronounced maxima. (d) Phase diagram of EuB$_6$ determined from the field sweeps of linear resistivity shown in (a) and weakly nonlinear transport in (c) in a semi-logarithmic plot. Solid and broken lines are fits to a linear dependence on the magnetic field (see also \cite{Das2012,ZhangPRL2009}).}
\label{figure2}
\end{figure}

In the paramagnetic region $T_{c_1} < T < T^\ast$, the MP are initially isolated and diluted in the electronic background 'sea'. 
When a magnetic field is applied, the MP increase in size \cite{Note2} until they form links, overlap (where at the same time their number may decrease) and finally merge. This leads to the dramatic decrease of the linear resistivity as a function of magnetic field, as shown in Fig.\,\ref{figure2}(a). The pronounced maximum observed in $R_{3\omega}(H)$ at different $T$ shown in Fig.\,\ref{figure2}(c) then reflects the field-induced percolation of the MP. Strikingly, a drop of the sample resistance precedes the peak in $R_{3\omega}$, see the phase diagram Fig.\,\ref{figure2}(d), similar to the situation when cooling in zero magnetic field through the ferromagnetic transition, where the resistivity first drops at $T_{c_1}$, accompanied by the onset of nonlinear transport, before the nonlinear signal peaks at $T_{c_2}$ . A striking observation is also that both the drop in the linear magnetoresistance and $R_{3\omega}^{\rm max}(H)$ show a linear temperature dependence, see Fig.\,\ref{figure2}(d), which extrapolates to the paramagnetic Curie temperature, similar to what is observed for the switching field in the Hall effect \cite{ZhangPRL2009}. This means that these phenomena occur at a single critical magnetization coinciding with the delocalization of the holes. In \cite{ZhangPRL2009} it has been estimated that this critical magnetization corresponds to about 10\,\% of the saturation moment of each Eu$^{2+}$. Remarably, it is suggested that approximately 15\,\% (which is of the same order) of the Eu moments order at the higher temperature transition $T_{c_1}$, and the remainder at the bulk ferromagnetic transition at $T_{c_2}$ \cite{SuellowPRB2000}.
In thin-film La$_{0.75}$Ca$_{0.25}$MnO$_3$ (LCMO) the metal-insulator transition is accompanied by the appearance of an intrinsic electrical nonlinearity, and the magnitude of $R_{3\omega}$ is suggested to describe the density of correlated Jahn-Teller polarons \cite{MoshnyagaPRB2009}. Since in our case, the magnitude of $R_{3\omega}(T,H)$ is related to the microgeometry of the phase-separated electronic system reflecting the density and arrangement of the MP, we can discuss our data in a qualitatively similar way. Accordingly, at high temperatures, slightly below $T^\ast$, only few MP are stable and large fields are necessary in order to achieve a substantial overlap. Upon lowering the temperature, obviously the number density of MP increases and smaller fields are needed to reach the percolation threshold. Also, the larger magnitude of the third-harmonic voltage signal indicates a higher degree of inhomogeneity due to the larger number of polaronic objects. 

\section{Conclusions}
In conclusion, the 'universal' observation of magnetic-order- and magnetic field-induced weak nonlinearity in the electronic transport provides evidence for magnetically-driven electronic phase separation in the CMR material EuB$_6$. Our data suggest that the charge delocalization is due to a percolation transition of magnetic polarons occurring at a single critical magnetization.

\begin{acknowledgments}
The work is supported by the Deutsche Forschungsgemeinschaft (DFG) through the Emmy Noether program.
\end{acknowledgments}


\begin{references}
\bibitem{DegiorgiPRL1997} L. Degiorgi, E. Felder,  H. R. Ott, J. L. Sarrao, and Z. Fisk, Phys. Rev. Lett. \textbf{79}, 5134 (1997).
\bibitem{CooleyPRB1997}J. C. Cooley, M. C. Aronson, J. L. Sarrao and Z. Fisk, Phys. Rev. B \textbf{56},14541 (1997).
\bibitem{MajumdarPRL1998} P. Majumdar and P. Littlewood, Phys. Rev. Lett. \textbf{81}, 1314 (1998).
\bibitem{TeresaNature1997} J. M. De Teresa, M. R. Ibarra, P. A. Igarabel, C. Ritter, C. Marquina, J. Blasco, J. Garcia, A. del Moral, and Z. Arnold, Nature \textbf{386}, 256 (1997).
\bibitem{NyhusPRB1997}P. Nyhus, S. Yoon, M. Kauffman, S. L. Cooper, Z. Fisk and J. Sarrao, Phys. Rev. B \textbf{56}, 2717 (1997).
\bibitem{SuellowPRB2000} S. S\"ullow, I. Prasad, M. C. Aronson, S. Bogdanovich, J. L. Sarrao, and Z. Fisk,  Phys. Rev. B \textbf{62}, 11626 (2000).



\bibitem{KasuyaRMP1968} T. Kasuya and A. Yanase, Rev. Mod. Phys. \textbf{40}, 684 (1968).
\bibitem{vonMolnarHandbook} S. von Moln\'{a}r and P. A. Stampe, Hand book of Magnetism and Advanced Magnetic Materials, Vol 5, Wiley Publishers.
\bibitem{CalderonPRB2004} M. J. Calder\'{o}n, L. G. L. Wegener, and P. B. Littlewood, Phys. Rev. B \textbf{70}, 092408 (2004).
\bibitem{ChatterjeePRB2004} J. Chatterjee, U. Yu, and B. Min, Phys. Rev. B \textbf{69}, 134423 (2004).
\bibitem{YuPRB2006} U. Yu and B. I. Min, Phys. Rev. B \textbf{74}, 094413 (2006).
\bibitem{BrooksPRB2004} M. L. Brooks, T. Lancaster, S. J. Blundell, W. Hayes, F. L. Pratt, and Z. Fisk, Phys. Rev. B \textbf{70}, 020401(R), (2004).

\bibitem{Das2012}P. Das {\it et al.}, to be published.

\bibitem{ZhangPRL2008} X. Zhang, S. von Moln\'{a}r, Z. Fisk, and P. Xiong, Phys. Rev. Lett.\textbf{100}, 167001 (2008).
\bibitem{ZhangPRL2009}X. Zhang, L. Yu, S. von Moln\'{a}r, Z. Fisk and P. Xiong, Phys. Rev. Lett.\textbf{103}, 106602 (2009).

\bibitem{MoshnyagaPRB2009} V. Moshnyaga, K. Gehrke, O. I. Lebedev, L. Sudheendra, A. Belenchuk, S. Raabe, O. Shapoval, J. Verbeeck, G. Van Tendeloo, and K. Samwar, Phys. Rev. B \textbf{79}, 134413 (2009).

\bibitem{FiskJAP1979}Z. Fisk \textit{et al.} J. Appl. Phys. \textbf{50}, 1911 (1979).
\bibitem{BergmanPRB1989} D. J. Bergman, Phys. Rev. B {\bf 39}, 4598 (1989).
\bibitem{Levy1994} O. Levy and D. J. Bergman, Phys. Rev. B {\bf 50}, 3652 (1994).
\bibitem{DubsonPRB1989} M. A. Dubson, Y. C. Hui, M. B. Weissman, and J. C. Garland, Phys. Rev. B \textbf{39}, 6807 (1989).

\bibitem{Bergman1992} D. J. Bergman and S. Stroud, {\em Physical properties of macroscopically inhomogeneous media}, in: Solid State Physics {\bf 46}, 147 -- 269 (1992).
\bibitem{YagilPRB1992} Y. Yagil and G. Deutscher, Phys. Rev. B. \textbf{46}, 16115 (1992).
\bibitem{SuellowPRB1998} S. S\"ullow, I. Prasad, M. C. Aronson, J. L. Sarrao, Z. Fisk, D. Hristova, A. H. Lacerda, M. F. Hundley, A. Vigilante and D. Gibbs, Phys. Rev. B \textbf{57}, 5860 (1998).

\bibitem{Kogan1996} Sh. Kogan, {\em Electronic noise and fluctuations in solids}, Cambridge University Press, New York (1996).

\bibitem{PaschenPRB2000} S. Paschen, D. Pushin, M. Schlatter, P. Volanthen, H. R. Ott, D. P. Young, and Z. Fisk, Phys. Rev. B \textbf{61}, 4174 (2000).




\bibitem{Note2} Spatial extent of magnetic polarons were found to be increased by application of external magnetic fields from the observation of ferromagnetic correlation length in manganites \cite{TeresaNature1997}. The data also suggest a reduction of the number of polarons due to application of magnetic field.


\end{references}
\end{document}